\documentclass[12pt]{article}

\usepackage{amsmath}
\usepackage{amssymb}
\usepackage{graphicx}
\usepackage{fullpage}

\newtheorem{theorem}{Theorem}
\newtheorem{definition}{Definition}
\newtheorem{lemma}{Lemma}

\newcommand{\IR}{\mathbb{R}}

\newcommand{\DG}{\mathord{\it DG}}
\newcommand{\VD}{\mathord{\it VD}}
\newcommand{\cl}{\mathord{\it cl}}
\newcommand{\Int}{\mathord{\it int}}
\newcommand{\arc}{\mathord{\it arc}}

\newcommand{\qed}{\rule{0.5em}{1.5ex}}
\newcommand{\fqed}{{\hfill~\qed}}
\newenvironment{proof}{{\noindent \bf Proof.}}
                      {{\hfill \fqed} \vspace{1em}}

\title{On the Stretch Factor of Convex Delaunay Graphs}
\author{
Prosenjit Bose\thanks{School of Computer Science,
    Carleton University, Ottawa, Ontario, K1S 5B6, Canada.
    These authors were supported by NSERC.}
\and 
Paz Carmi\footnotemark[1] 
\and   
S{\'e}bastien Collette\thanks{Charg{\'e} de recherches du F.R.S.-FNRS. 
Computer Science Department, Universit{\'e} Libre de Bruxelles, 
CP212, Bvd du Triomphe, 1050 Brussels, Belgium.} 
\and  
Michiel Smid\footnotemark[1] 
}
\date{\today} 

\begin{document} 
 
\maketitle

\begin{abstract}
Let $C$ be a compact and convex set in the plane that contains the origin 
in its interior, and let $S$ be a finite set of points in the plane. 
The Delaunay graph $\DG_C(S)$ of $S$ is defined to be the dual of the 
Voronoi diagram of $S$ with respect to the convex distance function 
defined by $C$. We prove that $\DG_C(S)$ is a $t$-spanner for $S$, 
for some constant $t$ that depends only on the shape of the set $C$. 
Thus, for any two points $p$ and $q$ in $S$, the graph $\DG_C(S)$ 
contains a path between $p$ and $q$ whose Euclidean length is at most 
$t$ times the Euclidean distance between $p$ and $q$.  
\end{abstract}

\section{Introduction}   \label{secintro}  
Let $S$ be a finite set of points in the plane and let $G$ be a graph 
with vertex set $S$, in which each edge $(p,q)$ has a weight equal to the 
Euclidean distance $|pq|$ between $p$ and $q$. For a real number 
$t \geq 1$, we say that $G$ is a $t$-\emph{spanner} for $S$, if for any 
two points $p$ and $q$ of $S$, there exists a path in $G$ between $p$ 
and $q$ whose Euclidean length is at most $t|pq|$. The smallest such 
$t$ is called the \emph{stretch factor} of $G$. The problem of 
constructing spanners has received much attention; see 
Narasimhan and Smid~\cite{ns-gsn-07} for an extensive overview. 

Spanners were introduced in computational geometry by 
Chew~\cite{c-tipga-86,c-tapga-89}, who proved the following two results. 
first, the $L_1$-Delaunay graph, i.e., the dual of the Voronoi diagram 
for the Manhattan metric, is a $\sqrt{10}$-spanner. Second, the Delaunay 
graph based on the convex distance function defined by an equilateral 
triangle, is a $2$-spanner. We remark that in both these results, the 
stretch factor is measured in the Euclidean metric. 
Chew also conjectured that the Delaunay graph based on the Euclidean 
metric, is a $t$-spanner, for some constant $t$. (If not all points of 
$S$ are on a line, and if no four points of $S$ are cocircular, then 
the Delaunay graph is the well-known Delaunay \emph{triangulation}.)  
This conjecture was proved by Dobkin \emph{et al.}~\cite{dfs-dgaag-90}, 
who showed that $t \leq \pi (1+\sqrt{5})/2$. The analysis was improved by 
Keil and Gutwin~\cite{kg-cgwac-92}, who showed that
$t \leq \frac{4 \pi \sqrt{3}}{9}$. 

In this paper, we unify these results by showing that the Delaunay graph 
based on any convex distance function has bounded stretch factor.  

Throughout this paper, we fix a compact and convex set $C$ in the plane. 
We assume that the origin is in the interior of $C$. A \emph{homothet} 
of $C$ is obtained by scaling $C$ with respect to the origin, followed 
by a translation. Thus, a homothet of $C$ can be written as 
\[ x + \lambda C = \{ x + \lambda z : z \in C \} , 
\] 
for some point $x$ in the plane and some real number $\lambda \geq 0$. 
We call $x$ the \emph{center} of the homothet $x + \lambda C$. 

For two points $x$ and $y$ in the plane, we define 
\[ d_C(x,y) := \min \{ \lambda \geq 0 : y \in x + \lambda C \} .
\] 
If $x \neq y$, then this definition is equivalent to the following: 
Consider the translate $x+C$ and the ray emanating from $x$ that 
contains $y$. Let $y'$ be the (unique) intersection between this ray 
and the boundary of $x+C$. Then 
\[ d_C(x,y) = |xy| / |xy'| . 
\] 
The function $d_C$ is called the \emph{convex distance function} 
associated with $C$. Clearly, we have $d_C(x,x)=0$ and $d_C(x,y) > 0$ 
for all points $x$ and $y$ with $x \neq y$. 
Chew and Drysdale~\cite{cd-vdbcd-85} showed that the triangle 
inequality $d_C(x,z) \leq d_C(x,y) + d_C(y,z)$ holds. In general, the 
function $d_C$ is not symmetric, i.e., $d_C(x,y)$ is not necessarily 
equal to $d_C(y,x)$. If $C$ is symmetric with respect to the origin, 
however, then $d_C$ is symmetric. 

\begin{figure}[t]
   \begin{center}
     \includegraphics[scale=0.7]{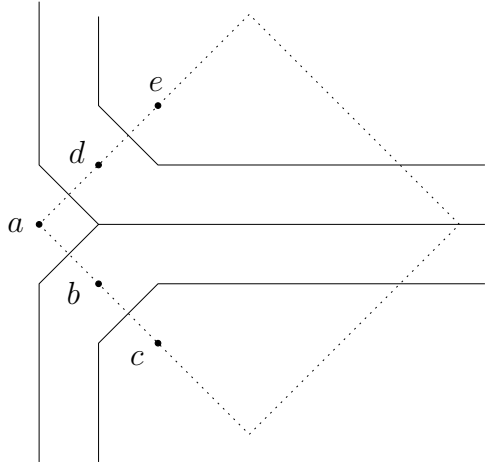}
     \caption{\sl The Voronoi diagram $\VD_C(S)$ for the set  
       $S = \{ a,b,c,d,e\}$. The set $C$ is the square as indicated 
   by the dotted figure; the origin is at the center of $C$.}
     \label{figexample}
   \end{center}
\end{figure}

Let $S$ be a finite set of points in the plane. For each point $p$ in 
$S$, we define  
\[ V'_C(p) := \{ x \in \IR^2 :  
                \mbox{ for all $q \in S$, $d_C(x,p) \leq d_C(x,q)$} 
             \} .
\]  
If $C$ is not strictly convex, then the set $V'_C(p)$ may consist of a 
closed region of positive area with an infinite ray attached to it. 
For example, in figure~\ref{figexample}, the set $V'_C(a)$ consists 
of the set of all points that are on or to the left of the leftmost 
zig-zag line, together with the infinite horizontal ray that is 
at the same height as the point $a$. Also, the intersection of two 
regions $V'_C(p)$ and $V'_C(q)$, where $p$ and $q$ are distinct points 
of $S$, may have a positive area. As a result, the collection $V'_C(p)$, 
where $p$ ranges over all points of $S$, does not necessarily give a 
subdivision of the plane in which the interior of each cell is associated 
with a unique point of $S$. In order to obtain such a subdivision, we 
follow the approach of Klein and Wood~\cite{kw-vdbgm-88} (see also 
Ma~\cite{m-bvdcd-00}): first, infinite rays attached to regions of 
positive area are not considered to be part of the region. Second, a 
point $x$ in $\IR^2$ that is in the interior of more than one region 
$V'_C(p)$ is assigned to the region of the lexicographically smallest 
point $p$ in $S$ for which $x \in V'_C(p)$. 

To formally define Voronoi cells, let $\prec$ denote the lexicographical 
ordering on the set of all points in the plane. Let 
$p_1 \prec p_2 \prec \ldots \prec p_n$ be the points of $S$, sorted 
according to this order. Then the \emph{Voronoi cells} $V_C(p_i)$ of the 
points of $S$ are defined as 
\[ V_C(p_1) := \cl ( \Int ( V'_C(p_1)) )   
\]
and, for $1 < i \leq n$,  
\[ V_C(p_i) := \cl \left( 
                 \Int \left( V'_C(p_i) \setminus 
                             \left( \bigcup_{j<i} V_C(p_j) \right) 
                      \right)  
                   \right) , 
\] 
where $\cl(X)$ and $\Int(X)$ denote the closure and the interior of the 
set $X \subseteq \IR^2$, respectively.  

Thus, in figure~\ref{figexample}, the Voronoi cell $V_C(a)$ consists only  
of the set of all points that are on or to the left of the leftmost 
zig-zag line; the infinite horizontal ray that is at the same height as 
the point $a$ is not part of this cell. 

The \emph{Voronoi diagram} $\VD_C(S)$ of $S$ with respect to $C$ is 
defined to be the collection of Voronoi cells $V_C(p)$, where $p$ ranges 
over all points of $S$. An example is given in figure~\ref{figexample}.  

As for the Euclidean case, the Voronoi diagram $\VD_C(S)$ induces 
Voronoi cells, Voronoi edges, and Voronoi vertices. Each point in the 
plane is either in the interior of a unique Voronoi cell, in the 
relative interior of a unique Voronoi edge, or a unique Voronoi vertex. 
Each Voronoi edge $e$ belongs only to the two Voronoi cells that contain 
$e$ on their boundaries.  
Observe that Voronoi cells are closed.  

The Delaunay graph is defined to be the dual of the Voronoi diagram: 

\begin{definition}  \label{defDG} 
       Let $S$ be a finite set of points in the plane. The 
       \emph{Delaunay graph} $\DG_C(S)$ of $S$ with respect to $C$ is 
       defined to be the dual of the Voronoi diagram $\VD_C(S)$. That is, 
       the vertex set of $\DG_C(S)$ is $S$ and two distinct vertices 
       $p$ and $q$ are connected by an edge in $\DG_C(S)$ if and only if 
       the Voronoi cells $V_C(p)$ and $V_C(q)$ share a Voronoi edge. 
\end{definition}

For example, the Delaunay graph $\DG_C(S)$ for the point set in 
figure~\ref{figexample} consists of the five edges $(a,b)$, $(a,d)$, 
$(b,c)$, $(b,d)$, and $(d,e)$. 

We consider the Delaunay graph $\DG_C(S)$ to be a geometric graph, which 
means that each edge $(p,q)$ is embedded as the closed line segment with 
endpoints $p$ and $q$. 

\begin{figure}[t]
   \begin{center}
     \includegraphics[scale=0.9]{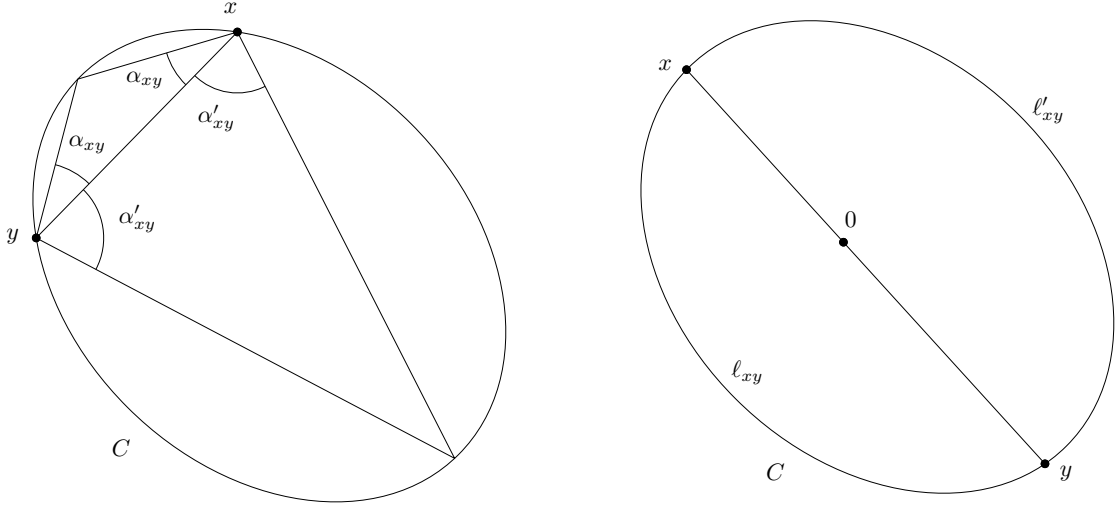}
     \caption{\sl The two parameters associated with $C$.}
     \label{figparam}
   \end{center}
\end{figure}

Before we can state the main result of this paper, we introduce two 
parameters whose values depend on the shape of the set $C$. 
Let $x$ and $y$ be two distinct points on the boundary $\partial C$ of 
$C$. These points partition $\partial C$ into two chains. For each of
these chains, there is an isosceles triangle with base $xy$ and 
whose third vertex is on the chain. Denote the base angles of these 
two triangles by $\alpha_{xy}$ and $\alpha'_{xy}$; see  
figure~\ref{figparam} (left). We define 
\[ \alpha_C := \min \{ \max( \alpha_{xy} , \alpha'_{xy} ) : 
                       x,y \in \partial C , x \neq y \} .
\] 

Consider again two distinct points $x$ and $y$ on $\partial C$, but now 
assume that $x$, $y$, and the origin are collinear. As before, $x$ and 
$y$ partition $\partial C$ into two chains. Let $\ell_{xy}$ and 
$\ell'_{xy}$ denote the lengths of these chains; see 
figure~\ref{figparam} (right). We define 
\[ \kappa_{C,0} := \max 
     \left\{ \frac{\max( \ell_{xy} , \ell'_{xy} )}{|xy|} :  
                x,y \in \partial C , x \neq y , 
                   \mbox{ and $x$, $y$, and $0$ are collinear} 
     \right\} .  
\] 
Clearly, the convex distance function $d_C$ and, therefore, the Voronoi 
diagram $\VD_C(S)$, depends on the location of the origin in the 
interior of $C$. Surprisingly, the Delaunay graph $\DG_C(S)$ does not 
depend on this location; see Ma \cite[Section~2.1.6]{m-bvdcd-00}. 
We define 
\[ \kappa_C := \min \left\{ \kappa_{C,0} : 
                       \mbox{ $0$ is in the interior of $C$}  
                    \right\} . 
\]  
In this paper, we will prove the following result: 

\begin{theorem}   \label{thmmain} 
       Let $C$ be a compact and convex set in the plane with a non-empty 
       interior, and let $S$ be a finite set of points in the plane. 
       The stretch factor of the Delaunay graph $\DG_C(S)$ is less than 
       or equal to 
       \[ t_C :=  
            \left\{ 
              \begin{array}{ll} 
                2 \kappa_C \cdot 
                      \max \left( \frac{3}{\sin (\alpha_C/2)} , \kappa_C 
                           \right) 
                       & \mbox{if $\DG_C(S)$ is a triangulation,} \\ 
                2 \kappa_C^2 \cdot 
                      \max \left( \frac{3}{\sin (\alpha_C/2)} , \kappa_C 
                           \right) 
                       & \mbox{otherwise.} \\ 
              \end{array} 
            \right. 
       \]  
       Thus, for any two points $p$ and $q$ in $S$, the graph 
       $\DG_C(S)$ contains a path between $p$ and $q$ whose Euclidean 
       length is at most $t_C$ times the Euclidean distance between 
       $p$ and $q$.  
\end{theorem} 

We emphasize that we do \emph{not} make any ``general position'' 
assumption; our proof of Theorem~\ref{thmmain} is valid for \emph{any} 
finite set of points in the plane. 

Throughout the rest of this paper, we assume that the origin is chosen 
in the interior of $C$ such that $\kappa_C = \kappa_{C,0}$. 

The rest of this paper is organized as follows.  
In Section~\ref{secprop}, we prove some basic properties of the 
Delaunay graph which are needed in the proof of Theorem~\ref{thmmain}. 
In particular, we give a formal proof of the fact that this graph is 
plane. Even though this fact seems to be well known, we have not been 
able to find a proof in the literature. 
Section~\ref{secSFDG} contains a proof of Theorem~\ref{thmmain}.  
This proof is obtained by showing that the Delaunay graph satisfies the 
``diamond property'' and a variant of the ``good polygon property'' 
of Das and Joseph~\cite{dj-wtacg-89}. The proof of the latter property 
is obtained by generalizing the analysis of 
Dobkin \emph{et al.}~\cite{dfs-dgaag-90} for the lengths of so-called 
one-sided paths.

\section{Some properties of the Delaunay graph}   
\label{secprop}  
Recall that in the Euclidean Delaunay graph, if two points $p$ and $q$ 
of $S$ are connected by an edge, then there exists a disk having $p$ and 
$q$ on its boundary that does not contain any point of $S$ in its 
interior. The next lemma generalizes this result to the Delaunay graph 
$\DG_C(S)$.  

\begin{lemma}   \label{lememptydisc} 
       Let $p$ and $q$ be two points of $S$ and assume that $(p,q)$ is an 
       edge in the Delaunay graph $\DG_C(S)$. Then, the following are 
       true. 
       \begin{enumerate} 
       \item The line segment between $p$ and $q$ does not contain any 
             point of $S \setminus \{p,q\}$.  
       \item For every point $x$ in 
             $V_C(p) \cap V_C(q)$, there exists a real number 
             $\lambda > 0$ such that 
             \begin{enumerate} 
             \item the homothet $x + \lambda C$ contains $p$ and $q$ on 
                   its boundary, and  
             \item the interior of $x + \lambda C$ does not contain any 
                   point of $S$.  
             \end{enumerate} 
       \end{enumerate} 
\end{lemma}  
\begin{proof} 
To prove the first claim, assume that the line segment between $p$ and 
$q$ contains a point of $S \setminus \{p,q\}$. Then it follows from 
Ma \cite[Lemma~2.1.4.2]{m-bvdcd-00} that $V_C(p) \cap V_C(q) = \emptyset$.  
Thus, the Voronoi cells of $p$ and $q$ do not share an edge and, 
therefore, $(p,q)$ is not an edge in the Delaunay graph. This is a 
contradiction. 

To prove the second claim, let $x$ be an arbitrary point in 
$V_C(p) \cap V_C(q)$. Then $d_C(x,p) = d_C(x,q)$ and 
$d_C(x,r) \geq d_C(x,p)$ for all $r \in S$. Thus, if we define 
$\lambda := d_C(x,p)$, then $\lambda > 0$, both $p$ and $q$ are on the 
boundary of the homothet $x + \lambda C$, and no point of $S$ is in the 
interior of this homothet.  
\end{proof}

As can be seen in figure~\ref{figexample}, Voronoi cells are, in general, 
not convex. They are, however, star-shaped: 

\begin{lemma}     \label{lemstarshaped}  
       Let $p$ be a point of $S$ and let $x$ be a point in the Voronoi 
       cell $V_C(p)$. Then the line segment $xp$ is completely contained 
       in $V_C(p)$. 
\end{lemma} 
\begin{proof}  
In \cite[Lemma~2.1.4.7]{m-bvdcd-00}, Ma shows that, if $x$ is in the 
interior of $V_C(p)$, then $xp$ is completely in the interior of 
$V_C(p)$. Clearly, this implies that $xp$ is in $V_C(p)$, if 
$x \in V_C(p)$ (i.e., $x$ is in the interior or on the boundary of this 
Voronoi cell). 
\end{proof} 

It is well known that the Euclidean Delaunay graph is a plane graph; 
see, for example, de Berg \emph{et al.}~\cite[page 189]{bkos-cgaa-97}. 
The following lemma states that this is true for the Delaunay graph 
$\DG_C(S)$ as well. 

\begin{lemma}    \label{lemplanar}
       The Delaunay graph $\DG_C(S)$ is a plane graph. 
\end{lemma}
\begin{proof} 
By the first claim in Lemma~\ref{lememptydisc}, $\DG_C(S)$ does not 
contain two distinct edges $(p,q)$ and $(p,r)$ that are collinear and 
overlap in a line segment of positive length. Again by the first claim 
in Lemma~\ref{lememptydisc}, $\DG_C(S)$ does not contain two distinct 
edges $(p,q)$ and $(r,s)$ such that $r$ is on the open line segment 
joining $p$ and $q$. 

It remains to show that $\DG_C(S)$ does not contain two edges $(p,q)$ and 
$(r,s)$ that cross properly. The proof is by contradiction. Thus, let 
$p$, $q$, $r$, and $s$ be four pairwise distinct points of $S$, no three 
of which are collinear, and assume that the line segments $(p,q)$ and 
$(r,s)$ are edges of $\DG_C(S)$ that have exactly one point in common. 

Since $(p,q)$ is an edge of $\DG_C(S)$, there exists a point $x$ in the 
relative interior of $V_C(p) \cap V_C(q)$. Thus, by the second claim in  
Lemma~\ref{lememptydisc}, there exists a real number $\lambda > 0$, 
such that the homothet $x + \lambda C$ contains $p$ and $q$ on its 
boundary and no point of $S$ is in the interior of this homothet. 
Observe that $x$ is in the interior of $x + \lambda C$. 
Let $D$ be a Euclidean disk centered at $x$ that is contained in the 
interior of $x + \lambda C$ and that is contained in 
$V_C(p) \cup V_C(q)$. We define $B$ to be the set of all 2-link 
polygonal chains $(p,z,q)$, with $z \in D$; see figure~\ref{figplane}. 
Observe that $B$ has a positive area. Since $V_C(p)$ and $V_c(q)$ are 
star-shaped (by Lemma~\ref{lemstarshaped}), we have 
$B \subseteq V_C(p) \cup V_C(q)$. Since $x + \lambda C$ is convex, we 
have $B \subseteq x + \lambda C$; in fact, the convex hull of $B$ is 
contained in $x + \lambda C$. Thus, neither $r$ nor $s$ is in the 
interior of the convex hull of $B$. Since $pq$ and $rs$ intersect in a 
point, the line segment $rs$ crosses the set $B$. 

\begin{figure}[t]
   \begin{center}
     \includegraphics[scale=0.7]{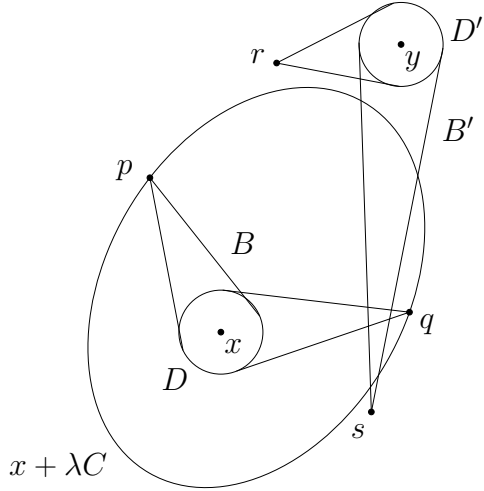}
     \caption{\sl Illustrating the proof of Lemma~\ref{lemplanar}.}
     \label{figplane}
   \end{center}
\end{figure}

By a symmetric argument, since $(r,s)$ is an edge of $\DG_C(S)$, there 
exist a point $y$ in the relative interior of $V_C(r) \cap V_C(s)$ and a 
real number $\mu > 0$, such that $y + \mu C$ contains $r$ and $s$ on its 
boundary and no point of $S$ is in the interior of this homothet. 
Let $D'$ be a Euclidean disk centered at $y$ that is contained in the 
interior of $y + \mu C$ and that is contained in $V_C(r) \cup V_C(s)$. 
We define $B'$ to be the set of all 2-link polygonal chains $(r,z,s)$, 
with $z \in D'$. The set $B'$ has a positive area, the line segment 
$pq$ crosses this set, $B' \subseteq V_C(r) \cup V_C(s)$, and neither $p$ 
nor $q$ is in the interior of the convex hull of $B'$. 

It follows that $B$ and $B'$ overlap in a region of positive area. 
Since $B \subseteq V_C(p) \cup V_C(q)$ and 
$B' \subseteq V_C(r) \cup V_C(s)$, however, the area of the intersection 
$B \cap B'$ is equal to zero. This is a contradiction. It follows that 
the edges $(p,q)$ and $(r,s)$ do not cross. 
\end{proof}

\section{The stretch factor of Delaunay graphs} 
\label{secSFDG} 

In this section, we will prove Theorem~\ref{thmmain}. first, we show that 
the Delaunay graph $\DG_C(S)$ satisfies the diamond property and a variant 
of the good polygon property of Das and Joseph~\cite{dj-wtacg-89}.  
According to the results of Das and Joseph, this immediately implies 
that the stretch factor of $\DG_C(S)$ is bounded. In fact, we will 
obtain an upper bound on the stretch factor which is better than the 
one that is implied by Das and Joseph's result. 

\subsection{The diamond property} 
\begin{figure}[t]
   \begin{center}
     \includegraphics[scale=0.9]{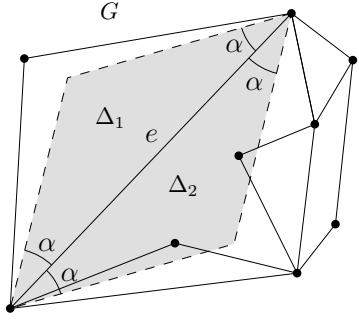}
     \caption{\sl The $\alpha$-diamond property.}
     \label{figdiamond}
   \end{center}
\end{figure}

Let $G$ be a plane graph with vertex set $S$ and let $\alpha$ be a real 
number with $0 < \alpha < \pi/2$. For any edge $e$ of $G$, let $\Delta_1$ 
and $\Delta_2$ be the two isosceles triangles with base $e$ and base
angle $\alpha$; see figure~\ref{figdiamond}. We say that $e$ 
satisfies the $\alpha$-\emph{diamond property}, if at least one of the 
triangles $\Delta_1$ and $\Delta_2$ does not contain any point of $S$ in 
its interior. The graph $G$ is said to satisfy the
$\alpha$-\emph{diamond property}, if every edge $e$ of $G$ satisfies
this property. 

\begin{figure}[t]
   \begin{center}
     \includegraphics[scale=0.9]{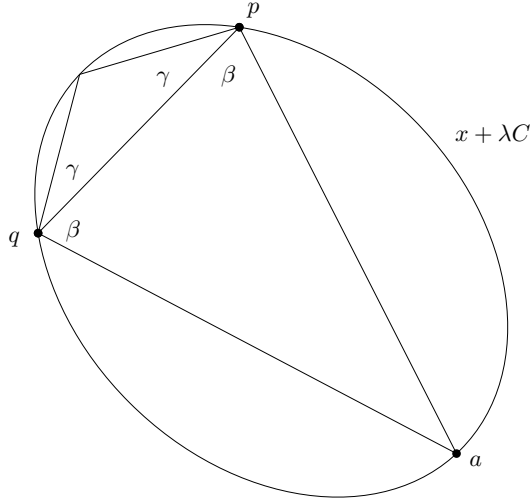}
     \caption{\sl Illustrating the proof of Lemma~\ref{lemdiamond}.}
     \label{figdiamondproof}
   \end{center}
\end{figure}

\begin{lemma}   \label{lemdiamond}
       Consider the value $\alpha_C$ that was defined in 
       Section~\ref{secintro}. The Delaunay graph $\DG_C(S)$ satisfies 
       the $\alpha_C$-diamond property. 
\end{lemma} 
\begin{proof}
Let $(p,q)$ be an arbitrary edge of $\DG_C(S)$ and let $x$ be any point 
in the relative interior of $V_C(p) \cap V_C(q)$. By 
Lemma~\ref{lememptydisc}, there exists a real number $\lambda > 0$ such 
that $p$ and $q$ are on the boundary of the homothet $x + \lambda C$ 
and no point of $S$ is in the interior of $x + \lambda C$. The points 
$p$ and $q$ partition $\partial (x + \lambda C)$ into two chains. For 
each of these chains, there is an isosceles triangle with base 
$pq$ and whose third vertex is on the chain. We denote the base angles 
of these two triangles by $\beta$ and $\gamma$; see 
figure~\ref{figdiamondproof}. We may assume without loss of generality 
that $\beta \geq \gamma$. Let $a$ denote the third vertex of the 
triangle with base angle $\beta$. If we translate $x + \lambda C$ so 
that $x$ coincides with the origin and scale the translated homothet 
by a factor of $1/\lambda$, then we obtain the set $C$. This 
translation and scaling does not change the angles $\beta$ and 
$\gamma$. Thus, using the notation of Section~\ref{secintro} (see 
also figure~\ref{figparam}), we have 
$\{ \beta , \gamma \} = \{ \alpha_{pq} , \alpha'_{pq} \}$. 
The definition of $\alpha_C$ then implies that 
\[ \alpha_C \leq \max ( \alpha_{pq} , \alpha'_{pq} ) = \beta .
\] 
Let $\Delta$ be the isosceles triangle with base $pq$ and base angle 
$\alpha_C$ such that $a$ and the third vertex of $\Delta$ are on the 
same side of $pq$. Then $\Delta$ is contained in the triangle with 
vertices $p$, $q$, and $a$. Since the latter triangle is contained in 
$x + \lambda C$, it does not contain any point of $S$ in its interior. 
Thus, $\Delta$ does not contain any point of $S$ in its interior. 
This proves that the edge $(p,q)$ satisfies the $\alpha_C$-diamond 
property. 
\end{proof} 
 
\subsection{The visible-pair spanner property} 
For a real number $\kappa \geq 1$, we say that the plane graph $G$ 
satisfies the \emph{strong visible-pair} $\kappa$-\emph{spanner property}, 
if the following is true: For every face $f$ of $G$, and for every two 
vertices $p$ and $q$ on the boundary of $f$, such that the open line 
segment joining $p$ and $q$ is completely in the interior of $f$, the 
graph $G$ contains a path between $p$ and $q$ having length at most 
$\kappa |pq|$. 
If for every face $f$ of $G$ and for every two vertices $p$ and $q$ on 
the boundary of $f$, such that the line segment $pq$ does not intersect 
the exterior of $f$, the graph $G$ contains a path between $p$ and $q$ 
having length at most $\kappa |pq|$, then we say that $G$ satisfies the 
\emph{visible-pair} $\kappa$-\emph{spanner property}. Observe that the 
former property implies the latter one. Also, observe that these 
properties are variants of the $\kappa$-\emph{good polygon} property of 
Das and Joseph~\cite{dj-wtacg-89}: The $\kappa$-good polygon property 
requires that $G$ contains a path between $p$ and $q$ that is along the 
boundary of $f$ and whose length is at most $\kappa |pq|$; in the 
(strong) visible-pair spanner property, the path is not required to be 
along the boundary of $f$. 

In this subsection, we will prove that the Delaunay graph $\DG_C(S)$ 
satisfies the visible-pair $\kappa_C$-spanner property, where $\kappa_C$ 
is as defined in Section~\ref{secintro}. This claim will be proved by 
generalizing results of Dobkin \emph{et al.}~\cite{dfs-dgaag-90} on 
so-called one-sided paths. 

Let $p$ and $q$ be two distinct points of $S$ and assume that $(p,q)$ 
is not an edge of the Delaunay graph $\DG_C(S)$. Consider the Voronoi 
diagram $\VD_C(S)$. We consider the sequence of points in $S$ whose 
Voronoi cells are visited when the line segment $pq$ is traversed from 
$p$ to $q$. If $pq$ does not contain any Voronoi vertex, then this 
sequence forms a path in $\DG_C(S)$ between $p$ and $q$. Since, in 
general, Voronoi cells are not convex, it may happen that this path 
contains duplicates. In order to avoid this, we define the sequence 
in the following way. 

In the rest of this section, we will refer to the line through $p$ and $q$ as the $X$-axis, and we will 
say that $p$ is to the \emph{left} of $q$. This implies a \emph{left-to-right} order on the $X$-axis, the notion of a point 
being \emph{above} or \emph{below} the X-axis, as well as the notions \emph{horizontal} and \emph{vertical}. (Thus, 
conceptually, we rotate and translate all points of $S$ , the set $C$, the Voronoi diagram $\VD_C(S)$, 
and the $\DG_C(S)$, such that $p$ and $q$ are on a horizontal line and $p$ is to the left 
of $q$. Observe that $\VD_C(S)$ is still defined based on the lexicographical order of the points of S 
before this rotation and translation.) In the following, we consider the (horizontal) line segment 
$pq$. If this segment contains a Voronoi vertex, then we imagine moving $pq$ vertically upwards by 
an infinitesimal amount. Thus, we may 
assume that $pq$ does not contain any Voronoi vertex of the 
(rotated and translated) Voronoi diagram $\VD_C(S)$.  
 
The first point in the sequence is $p_0 := p$. We define 
$x_1 \in \IR^2$ to be the point on the line segment $pq$ such 
that $x_1 \in V_C(p_0)$ and $x_1$ is closest to $q$. 

Let $i \geq 1$ and assume that the points $p_0,p_1,\ldots,p_{i-1}$ of 
$S$ and the points $x_1,\ldots,x_i$ in $\IR^2$ have already been 
defined, where $x_i$ is the point on the line segment $pq$ such that 
$x_i \in V_C(p_{i-1})$ and $x_i$ is closest to $q$. If $p_{i-1} = q$, 
then the construction is completed. Otherwise, observe that $x_i$ is in 
the relative interior of a Voronoi edge. We define $p_i$ to be the point 
of $S \setminus \{ p_{i-1} \}$ whose Voronoi cell contains $x_i$ on its 
boundary, and define $x_{i+1}$ to be the point on the line segment 
$pq$ such that $x_{i+1} \in V_C(p_i)$ and $x_{i+1}$ is closest to $q$. 

Let $p=p_0,p_1,\ldots,p_k =q$ be the sequence of points in $S$ obtained 
in this way. By construction, these $k+1$ points are pairwise distinct 
and for each $i$ with $1 \leq i \leq k$, the Voronoi cells 
$V_C(p_{i-1})$ and $V_C(p_i)$ share an edge. Therefore, 
by definition, $(p_{i-1},p_i)$ is an edge in $\DG_C(S)$. Thus, 
$p=p_0,p_1,\ldots,p_k =q$ defines a path in $\DG_C(S)$ between $p$ and 
$q$. We call this path the \emph{direct path} between $p$ and $q$. If 
all points $p_1,p_2,\ldots,p_{k-1}$ are strictly on one side of the line 
through $p$ and $q$, then we say that the direct path is 
\emph{one-sided}.  

We will show in Lemma~\ref{lemonesided} that the length of a one-sided 
path is at most $\kappa_C |pq|$. The proof of this lemma uses a 
geometric property which we prove first.   

Let $C'$ be a homothet of $C$ whose center is on the $X$-axis, and let 
$x$ and $y$ be two points on the boundary of $C'$ that are on or above 
the $X$-axis. The points $x$ and $y$ partition the boundary of $C'$ 
into two chains. One of these chains is completely on or above the 
$X$-axis; we denote this chain by $\arc(x,y;C')$. The length of this 
chain is denoted by $|\arc(x,y;C')|$. 

For two points $x$ and $y$ on the $X$-axis, we write $x <_X y$ if 
$x$ is strictly to the left of $y$, and we write $x \leq_X y$ if 
$x=y$ or $x <_X y$. 

\begin{figure}[t]
   \begin{center}
     \includegraphics[scale=0.5]{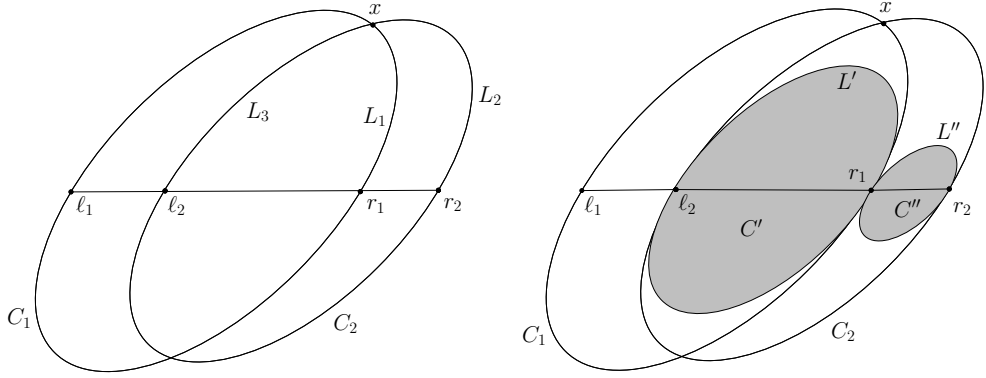}
     \caption{\sl Illustrating the proof of Lemma~\ref{lemtechnical}.}
     \label{figtechnical}
   \end{center}
\end{figure}

We now state the geometric property, which is illustrated in 
figure~\ref{figtechnical}. Recall the value $\kappa_C$ that was 
defined in Section~\ref{secintro}. 

\begin{lemma}   \label{lemtechnical} 
       Let $C_1 = y_1 + \lambda_1 C$ and $C_2 = y_2 + \lambda_2 C$ be 
       two homothets of $C$ whose centers $y_1$ and $y_2$ are on the 
       $X$-axis. Assume that $\lambda_1 > 0$, $\lambda_2 > 0$, and 
       $y_1 <_X y_2$. For $i=1,2$, let $\ell_i$ and $r_i$ be the leftmost 
       and rightmost points of $C_i$ on the $X$-axis, respectively. 
       Assume that $r_1 \leq_X r_2$ and $\ell_1 \leq_X \ell_2 <_X r_1$. 
       Let $x$ be a point that is on the boundaries of both $C_1$ and 
       $C_2$ and on or above the $X$-axis. Let 
       $L_1 = | \arc(x,r_1;C_1)|$ and $L_2 = | \arc(x,r_2;C_2)|$. Then 
       \[ L_2 \leq L_1 + \kappa_C |r_1 r_2| . 
       \] 
\end{lemma} 
\begin{proof} 
We define $L_3 = |\arc(\ell_2,x;C_2)|$. 
Let $C'$ be the homothet of $C$ whose center is on the $X$-axis such that 
the intersection between $C'$ and the $X$-axis is equal to the line 
segment $\ell_2 r_1$, and let $L' = | \arc(\ell_2,r_1;C')|$; 
see figure~\ref{figtechnical}. Observe that, for 
$\lambda := |\ell_2 r_1|/|\ell_2 r_2|$, $C'$ is obtained from $C_2$ by 
a scaling by a factor of $\lambda$. Thus, since 
$| \arc(\ell_2,r_2;C_2)| = L_2 + L_3$, we have 
\[ L' = \lambda ( L_2 + L_3 ) .
\] 
Let $C''$ be the homothet of $C$ whose center is on the $X$-axis such 
that the intersection between $C''$ and the $X$-axis is equal to the 
line segment $r_1 r_2$, and let $L'' = | \arc(r_1,r_2;C'')|$. Since 
$C''$ is obtained from $C_2$ by a scaling by a factor of $1-\lambda$, 
we have 
\[ L'' = (1-\lambda) ( L_2 + L_3 ) .
\] 
Thus, we have 
\[ L' + L'' = L_2 + L_3 .
\] 
By convexity, we have $C' \subseteq C_1 \cap C_2$. Then it follows, 
again from convexity (see Benson~\cite[page 42]{b-egc-66}), that 
\[ L' \leq L_1 + L_3 .
\] 
Thus, we have 
\[ L_2 + L_3 = L' + L'' \leq L_1 + L_3 + L'' ,
\]
which implies that 
\[ L_2 \leq L_1 + L'' .
\]  
Since, by the definition of $\kappa_C$, $L'' \leq \kappa_C |r_1 r_2|$,
the proof is complete. 
\end{proof} 

We are now ready to prove an upper bound on the length of a one-sided 
path. 

\begin{lemma}  \label{lemonesided}  
       If the direct path between $p$ and $q$ is one-sided, then its 
       length is at most $\kappa_C |pq|$.  
\end{lemma} 
\begin{proof} 
As above, we assume that $p$ and $q$ are on the $X$-axis and that $p$ is 
to the left of $q$. Consider the direct path $p=p_0,p_1,\ldots,p_k =q$ 
in $\DG_C(S)$ and the sequence $x_1,x_2,\ldots,x_k$, as defined above. 
Since the direct path is one-sided, we may assume without loss of 
generality that the points $p_1,p_2,\ldots,p_{k-1}$ are strictly above 
the $X$-axis. We have to show that 
\begin{equation}    \label{eqtoshow}  
   \sum_{i=1}^k | p_{i-1} p_i | \leq \kappa_C |pq| .
\end{equation} 
 
Recall that, for each $i$ with $1 \leq i \leq k$, $x_i$ is in the 
relative interior of $V_C(p_{i-1}) \cap V_C(p_i)$ and $x_i$ is on the 
line segment $pq$. Therefore, by Lemma~\ref{lememptydisc}, if we 
define $\lambda_i := d_C(x_i,p_{i-1})$ (which is equal to 
$d_C(x_i,p_i)$), then the homothet $C_i := x_i + \lambda_i C$ contains 
$p_{i-1}$ and $p_i$ on its boundary and no point of $S$ is in its 
interior. 

For each $i$ with $1 \leq i \leq k$, let $\ell_i$ and $r_i$ be the 
leftmost and rightmost points of $C_i$ that are on the $X$-axis, 
respectively. We will prove that for each $j$ with $1 \leq j \leq k$, 
\begin{equation}    \label{eqwillshow}  
   \sum_{i=1}^{j-1} | p_{i-1} p_i | + | \arc(p_{j-1},r_j;C_j) | 
                 \leq \kappa_C |p r_j| .
\end{equation} 
For $j=k$, inequality (\ref{eqwillshow}) implies (\ref{eqtoshow}), 
because $r_k = p_k = q$. 

Before we prove (\ref{eqwillshow}), we show that 
$\ell_1 \leq_X \ell_2 \leq_X \ldots \leq_X \ell_k$. 
Observe that $x_1 <_X x_2 <_X \ldots <_X x_k$. 
Assume that there is an index $i$ such that $\ell_i <_X \ell_{i-1}$.  
Since $\ell_i <_X \ell_{i-1} <_X x_{i-1} <_X x_i$, it follows that 
$\lambda_{i-1} < \lambda_i$. If $r_{i-1} <_X r_i$, then 
$\ell_i <_X \ell_{i-1} <_X r_{i-1} <_X r_i$ and, therefore, 
$C_{i-1}$ is completely contained in the interior of $C_i$. 
This is a contradiction, because $p_{i-1}$ is on the boundary of 
$C_{i-1}$, but no point of $S$ is in the interior of $C_i$. 
Thus, we have $r_i \leq_X r_{i-1}$. Since 
$x_{i-1} <_X x_i <_X r_i \leq_X r_{i-1}$, we have 
$\lambda_{i-1} > \lambda_i$, which is a contradiction. 

Thus, we have shown that 
$\ell_1 \leq_X \ell_2 \leq_X \ldots \leq_X \ell_k$. 
By a symmetric argument, it follows that 
$r_1 \leq_X r_2 \leq_X \ldots \leq_X r_k$. 

Now we are ready to prove (\ref{eqwillshow}). The proof is by induction 
on $j$. For the base case, i.e., when $j=1$, we have to show that 
\[ | \arc(p_0,r_1;C_1) | \leq \kappa_C |p r_1| .
\] 
Since $p_0 = p = \ell_1$, this inequality follows from the definition 
of $\kappa_C$.  

Let $1 \leq j < k$ and assume that (\ref{eqwillshow}) holds for $j$. 
We have to show that (\ref{eqwillshow}) holds for $j+1$, i.e.,  
\begin{equation}  \label{eqtoshow2}  
   \sum_{i=1}^j | p_{i-1} p_i | + | \arc(p_j,r_{j+1};C_{j+1}) | 
                 \leq \kappa_C |p r_{j+1}| .
\end{equation}   
By the induction hypothesis, we have 
\begin{eqnarray*} 
  \lefteqn{\sum_{i=1}^j | p_{i-1} p_i | + 
           | \arc(p_j,r_{j+1};C_{j+1}) |} \\ 
  & = & \sum_{i=1}^{j-1} | p_{i-1} p_i | + | p_{j-1} p_j | + 
             | \arc(p_j,r_{j+1};C_{j+1}) |  \\ 
  & \leq & 
    \kappa_C |p r_j| - | \arc(p_{j-1},r_j;C_j) | + | p_{j-1} p_j | + 
             | \arc(p_j,r_{j+1};C_{j+1}) |  \\ 
  & = & 
    \kappa_C \left( |p r_{j+1}| - |r_j r_{j+1}| \right) - 
    | \arc(p_{j-1},r_j;C_j) | + | p_{j-1} p_j | + 
             | \arc(p_j,r_{j+1};C_{j+1}) | .
\end{eqnarray*} 
Thus, (\ref{eqtoshow2}) holds if we can show that 
\begin{equation}  \label{eqtoshow3}  
       | p_{j-1} p_j | + | \arc(p_j,r_{j+1};C_{j+1}) | \leq 
       | \arc(p_{j-1},r_j;C_j) | + \kappa_C |r_j r_{j+1}| . 
\end{equation}   
We distinguish two cases. 

\vspace{0.5em} 

\noindent 
{\bf Case 1:} $r_j \leq_X \ell_{j+1}$. 

By the triangle inequality, we have 
\[ | p_{j-1} p_j | \leq | p_{j-1} r_j | + | r_j \ell_{j+1} | 
                          + | \ell_{j+1} p_j | .
\] 
Since $p_j$ is on the boundary of $C_{j+1}$ and strictly above the 
$X$-axis, we have  
\begin{eqnarray*} 
  | \ell_{j+1} p_j | + | \arc(p_j,r_{j+1};C_{j+1}) | & \leq & 
     | \arc(\ell_{j+1},p_j;C_{j+1}) | + 
     | \arc(p_j,r_{j+1};C_{j+1}) | \\ 
   & = & | \arc(\ell_{j+1},r_{j+1};C_{j+1}) | \\ 
   & \leq & \kappa_C | \ell_{j+1} r_{j+1} | . 
\end{eqnarray*} 
It follows that 
\begin{eqnarray*} 
   | p_{j-1} p_j | + | \arc(p_j,r_{j+1};C_{j+1}) | & \leq &  
   | p_{j-1} r_j | + | r_j \ell_{j+1} | + 
           \kappa_C | \ell_{j+1} r_{j+1} | \\ 
      & \leq & | \arc(p_{j-1},r_j;C_j) | + 
               \kappa_C | r_j \ell_{j+1} | + 
               \kappa_C | \ell_{j+1} r_{j+1} |  \\ 
      & = & 
       | \arc(p_{j-1},r_j;C_j) | + \kappa_C |r_j r_{j+1}| . 
\end{eqnarray*} 
Thus, (\ref{eqtoshow3}) holds.   

\vspace{0.5em} 

\noindent 
{\bf Case 2:} $\ell_{j+1} <_X r_j$. 

Since $p_j$ is on the boundaries of both $C_j$ and $C_{j+1}$ and 
strictly above the $X$-axis, we can apply Lemma~\ref{lemtechnical} 
with $x= p_j$ and obtain 
\[ | \arc(p_j,r_{j+1};C_{j+1})| \leq 
   | \arc(p_j,r_j;C_j)| + \kappa_C |r_j r_{j+1}| . 
\] 
Thus, 
\[ | p_{j-1} p_j | + | \arc(p_j,r_{j+1};C_{j+1}) | \leq 
   | p_{j-1} p_j | + | \arc(p_j,r_j;C_j)| + \kappa_C |r_j r_{j+1}| . 
\] 
We claim that $p_j \in \arc(p_{j-1},r_j,C_j)$. Assuming this is true, 
it follows that 
\begin{eqnarray*} 
  | p_{j-1} p_j | + | \arc(p_j,r_{j+1};C_{j+1}) |  
   & \leq & 
   | \arc(p_{j-1},p_j;C_j) | + | \arc(p_j,r_j;C_j)| + 
              \kappa_C |r_j r_{j+1}| \\ 
   & = & 
   | \arc(p_{j-1},r_j;C_j) | + \kappa_C |r_j r_{j+1}| , 
\end{eqnarray*} 
i.e., (\ref{eqtoshow3}) holds.   

It remains to prove that $p_j \in \arc(p_{j-1},r_j,C_j)$.  
Since $p_0 = \ell_0$ and $p_1$ is strictly above the $X$-axis, 
this is true for $j=1$. Assume that $2 \leq j < k$ and 
$p_j \not\in \arc(p_{j-1},r_j,C_j)$. Then, since $p_j$ is strictly 
above the $X$-axis, $p_{j-1}$ is in the relative interior of 
$\arc(p_j,r_j,C_j)$. 

By the definition of the point $x_j$, there is a point $y$ on the 
$X$-axis such that $y <_X x_j$ and the line segment $y x_j$ is contained 
in the Voronoi cell $V_C(p_{j-1})$. By Lemma~\ref{lemstarshaped}, the 
triangle $\Delta$ with vertices $p_{j-1}$, $y$, and $x_j$ is contained 
in $V_C(p_{j-1})$. 

Again by the definition of the point $x_j$, there is a point $z$ on the 
$X$-axis such that $x_j <_X z$ and the line segment $x_j z$ is contained 
in the Voronoi cell $V_C(p_j)$. By Lemma~\ref{lemstarshaped}, the triangle 
$\Delta'$ with vertices $p_j$, $x_j$, and $z$ is contained in $V_C(p_j)$. 

Since $p_{j-1}$ and $p_j$ are strictly above the $X$-axis and since 
$p_{j-1}$ is in the relative interior of $\arc(p_j,r_j,C_j)$, the 
intersection of $\Delta$ and $\Delta'$ has a positive area and is 
contained in the intersection of $V_C(p_{j-1})$ and $V_C(p_j)$. This is a 
contradiction, because the area of the intersection of any two Voronoi 
cells is zero. 
\end{proof} 

We are now ready to prove that the Delaunay graph satisfies the 
visible-pair spanner property: 

\begin{lemma}    \label{lemVP}
       The Delaunay graph $\DG_C(S)$ satisfies the visible-pair 
       $\kappa_C$-spanner property. 
\end{lemma} 
\begin{proof} 
Recall from Lemma~\ref{lemplanar}, that the graph $\DG_C(S)$ is plane. 
It suffices to prove that $\DG_C(S)$ satisfies the strong visible-pair 
$\kappa_C$-spanner property. Let $f$ be a face of $G$ and let $p$ and 
$q$ be two vertices on $f$ such that the open line segment between $p$ 
and $q$ is contained in the interior of $f$. We have to show that there 
is a path in $\DG_C(S)$ between $p$ and $q$ whose length is at most 
$\kappa_C |pq|$. 

As before, we assume that $p$ and $q$ are on the $X$-axis and that $p$ is 
to the left of $q$. Consider the direct path $p=p_0,p_1,\ldots,p_k =q$ 
in $\DG_C(S)$ and the sequence $x_1,x_2,\ldots,x_k$, as defined in the 
beginning of this section. We will show that the direct path is 
one-sided. The lemma then follows from Lemma~\ref{lemonesided}. 

Since the open line segment between $p$ and $q$ is in the interior of 
$f$, none of the points $p_1,\ldots,p_{k-1}$ is on the closed line 
segment $pq$. Assume that for some $i$ with $1 \leq i < k$, $p_i$ is on the $X$-axis. 
Then $p_i$ is either strictly to the left of $p$ or strictly to the right 
of $q$. We may assume without loss of generality that $p_i$ is strictly 
to the right of $q$. Consider the point $x_i$ and the homothet 
$C_i = x_i + \lambda_i C$ as in the proof of Lemma~\ref{lemonesided}. 
Since $x_i$ is on $pq$ and in the interior of $C_i$, and since $p_i$ 
is on the boundary of $C_i$, it follows from convexity that $q$ is in 
the interior of $C_i$, which is a contradiction. 
Thus we have shown that none of the points $p_1,\ldots,p_{k-1}$ is on 
the $X$-axis. 
 
Assume that the direct path is not one-sided. Then there is an edge 
$(p_{i-1},p_i)$ on this path such that one of $p_{i-1}$ and $p_i$ is 
strictly below the $X$-axis and the other point is strictly above the 
$X$-axis. Let $z$ be the intersection between $p_{i-1} p_i$ and the 
$X$-axis. By assumption, $z$ is not on the open line segment joining 
$p$ and $q$, and by Lemma~\ref{lememptydisc}, $z \neq p$ and $z \neq q$. 
Thus, $z$ is either strictly to the left of $p$ or strictly to the right 
of $q$. We may assume without loss of generality that $z$ is strictly 
to the right of $q$. Consider again the point $x_i$ and the homothet 
$C_i = x_i + \lambda_i C$ as in the proof of Lemma~\ref{lemonesided}. 
This homothet contains the points $x_i$, $p_{i-1}$ and $p_i$. Thus, by 
convexity, $C_i$ contains the triangle with vertices $x_i$, $p_{i-1}$, 
and $p_i$. Since $q$ is in the interior of this triangle, it follows 
that $q$ is in the interior of $C_i$, which is a contradiction. 
\end{proof}

\subsection{The proof of Theorem~\ref{thmmain}} 
Das and Joseph~\cite{dj-wtacg-89} have shown that any plane graph 
satisfying the diamond property and the good polygon property has a 
bounded stretch factor. The analysis of the stretch factor was slightly 
improved by Bose \emph{et al.}~\cite{bls-ogds-07}. A close inspection 
of the proof in~\cite{bls-ogds-07} shows that the following holds: 
Let $G$ be a geometric graph with the following four properties: 
\begin{enumerate}
\item $G$ is plane. 
\item $G$ satisfies the $\alpha$-diamond property. 
\item The stretch factor of any one-sided path in $G$ is at most 
      $\kappa$. 
\item $G$ satisfies the visible-pair $\kappa'$-spanner property. 
\end{enumerate} 
Then, $G$ is a $t$-spanner for 
\[ t = 2 \kappa \kappa' \cdot 
       \max \left( \frac{3}{\sin (\alpha/2)} , \kappa \right) . 
\]  
We have shown that the Delaunay graph $\DG_C(S)$ satisfies all these 
properties: By Lemma~\ref{lemplanar}, $\DG_C(S)$ is plane. By 
Lemma~\ref{lemdiamond}, $\DG_C(S)$ satisfies the $\alpha_C$-diamond 
property. By Lemma~\ref{lemonesided}, the stretch factor of any one-sided 
path in $\DG_C(S)$ is at most $\kappa_C$. By Lemma~\ref{lemVP}, 
$\DG_C(S)$ satisfies the visible-pair $\kappa_C$-spanner property.  
If $\DG_C(S)$ is a triangulation, then obviously, $\DG_C(S)$ satisfies 
the visible-pair $1$-spanner property. Therefore, we have completed the 
proof of Theorem~\ref{thmmain}.

\section{Concluding remarks} 
We have considered the Delaunay graph $\DG_C(S)$, where $C$ is a compact
and convex set with a non-empty interior and $S$ is a finite set of
points in the plane. We have shown that the (Euclidean) stretch factor
of $\DG_C(S)$ is bounded from above by a function of two parameters
$\alpha_C$ and $\kappa_C$ that are determined only by the shape of $C$.
Roughly speaking, these two parameters give a measure of the ``fatness''
of the set $C$.

Our analysis provides the first generic bound valid for any compact and 
convex set $C$. In all previous works, only special examples of such 
sets $C$ were considered. Furthermore, our approach does not make any
``general position'' assumption about the point set $S$, while most
related works on Delaunay graphs do not consider the case when four 
points are cocircular.

Note that for the Euclidean Delaunay triangulation (i.e., when the set
$C$ is the disk of radius one, and with no four cocircular points), we 
have $\alpha_C = \pi/4$ and $\kappa_C = \pi/2$, and we derive an upper 
bound on the stretch factor of $\frac{3 \pi}{\sin (\pi/8)} \approx 24.6$.

Observe that this is much worse than the currently best known upper bound 
(as proved by Keil and Gutwin~\cite{kg-cgwac-92}), which is 
$\frac{4 \pi \sqrt{3}}{9} \approx 2.42$.
We leave open the problem of improving our upper bound. In particular,
is it possible to generalize the techniques of
Dobkin \emph{et al.}~\cite{dfs-dgaag-90} and
Keil and Gutwin~\cite{kg-cgwac-92}, from the Euclidean metric to
an arbitrary convex distance function?

\bibliographystyle{plain}
\bibliography{delaunayconvex}

\end{document}